\documentclass[amssymb,prl,aps,floatfix,floats,preprintnumbers,showpacs,showkeys,twocolumn,superscriptaddress,groupedaddress]{revtex4-1}
\usepackage{dcolumn}
\usepackage{bm}
\usepackage{amssymb}
\usepackage{rcs}
\usepackage{color}
\usepackage{graphicx}
\usepackage{graphics}
\usepackage[normalem]{ulem}
\usepackage{chngcntr}
\usepackage[thinspace,thinqspace]{SIunits}
\usepackage{natbib}
\usepackage{amsmath}
\newcommand{\muB}{\ensuremath{\mu_{B}}}
\newcommand{\mueff}{\ensuremath{\mu_{eff}}}

\newcommand{\CIG}{CeIr$_{3}$Ge$_{7}$}
\newcommand{\CRG}{CeRh$_{3}$Ge$_{7}$}
\newcommand{\CCA}{CeCd$_{3}$As$_{3}$}

\newcommand{\CPAG}{CePdAl$_{4}$Ge$_{2}$}
\begin{document}
\preprint{APS/123-QED}
\title{Crystalline electric field of Ce in trigonal symmetry: \CIG\ as a model case}
\author{J. Banda$^{1}$, B. K. Rai$^{2}$, H. Rosner$^{1}$, E. Morosan$^{2}$, C. Geibel$^{1}$ and M. Brando$^{1}$}
\affiliation{$^{1}$Max Planck Institute for Chemical Physics of Solids, N\"othnitzer Str. 40, 01187 Dresden, Germany\\ $^{2}$Department of Physics and Astronomy, Rice University, Houston, Texas 77005, USA}
\date{\today}
\begin{abstract}
The crystalline electric field (CEF) of Ce$^{3+}$ in trigonal symmetry has recently become of some relevance, for instance, in the search of frustrated magnetic systems. Fortunately, it is one of the CEF case in which a manageable analytic solution can be obtained. Here, we present this solution for the general case, and use this result to determine the CEF scheme of the new compound \CIG\ with the help of $T$-dependent susceptibility and isothermal magnetization measurements. The resulting CEF parameters $B_{2}^{0} = 34.4$\,K, $B_{4}^{0} = 0.82$\,K and $B_{4}^{3} = 67.3$\,K correspond to an exceptional large CEF splittings of the first and second excited levels, 374\,K and 1398\,K, and a large mixing between the $\left|\pm\frac{5}{2}\right>$ and the $\left|\mp\frac{1}{2}\right>$ states. This indicates a very strong easy plane anisotropy with an unusual small $c$-axis moment. Using the same general expressions, we show that the properties of the recently reported system \CCA\ can also be described by a similar CEF scheme, providing a much simpler explanation for its magnetic properties than the initial proposal. Moreover, a similar strong easy plane anisotropy has also been reported for the two compounds CeAuSn and \CPAG, indicating that the CEF scheme elaborated here for \CIG\ corresponds to an exemplary case for Ce$^{3+}$ in trigonal symmetry.
\end{abstract}
\pacs{71.70.Ch,75.10.Dg.}
\keywords{Crystalline electric field, \CIG.}
\maketitle
\section{Introduction}
Cerium-based intermetallic compounds have been the subject of intensive research during the past decades. This is due to the variety of unconventional and remarkable properties that has been identified in this class of materials, as for instance heavy-fermion superconductivity~\cite{Steglich1979,Petrovic2001}, multipolar order~\cite{Doenni2000,Sera2001}, Kondo insulator ground state~\cite{Jaime2000,Bruening2010} or non-Fermi-liquid behaviour associated with the presence of a quantum critical point~\cite{Petrovic2001,Loehneysen1994}. More recently, systems with geometrical frustrated structures have been considered for the search of spin liquid ground states~\cite{Doenni1996,Lucas2017,Sibille2015}.

The central role is played by the valence instability of the cerium $4f$-electron. In the Ce$^{3+}$ valence state cerium has a local moment with $S = 1/2$, $L = 3$ and total angular momentum $J = 5/2$, according to Hund's rules. Ce$^{4+}$ is non-magnetic. In metals there are then three relevant energy scales which determine the ground state of the system: The crystalline electric field (CEF), the distance of the $4f$-electron energy level ($E_{f}$) from the Fermi level ($E_{F}$), $\Delta = E_{F} -  E_{f}$, and the hybridization width $W = \pi N(E_{F})V_{sf}^{2}$, with $V_{sf}$ the hybridization strength between the $4f$ and the conduction electrons and $N(E_{F})$ the density of states at the Fermi level (see, e.g., Ref.~\cite{Bauer1991} and references therein). Three scenarios should therefore be considered depending on the relative magnitudes of these energy scales: i) For $W \geq \Delta$, the system shows intermediate-valence behaviour characterized by a nearly $T$-independent susceptibility at low $T$ and Fermi liquid ground state with weakly renormalized quasi particles~\cite{Lawrence1981}; ii) For $W < \Delta$, the Kondo effect is present and the system forms a singlet ground state with heavy renormalized quasi particles (heavy fermions)~\cite{Stewart1984}; iii) For $W << \Delta$, the system shows a stable valence state and long-range magnetic ordering which is essentially controlled by the RKKY (Ruderman-Kittel-Kasuya-Yosida) interaction. In Ce- and Yb-based systems the ordering temperatures, and thus exchange interactions, are weak ($< 10$\,K) compared to the other rare-earth-based systems, because of the tiny de Gennes factor. The ordering has been found to be usually antiferromagnetic (AFM), but recently several ferromagnetic systems were discovered~\cite{Brando2016}. Depending on the CEF, Ce-based systems may also exhibit multipolar order: For instance, in the high-symmetry cubic structure the CEF yields to a quartet $\Gamma_{8}$ ground state which allows quadrupolar and AFM order like in Ce$_{3}$Pd$_{20}$Ge$_{6}$~\cite{Doenni2000,Kitagawa1996} or Ce$_{1-x}$La$_{x}$B$_{6}$~\cite{Fujita1980,Jang2017}.

In recent past years several Ce-based systems were studied with a trigonal symmetry for the Ce atoms, which is prone to frustration~\cite{Adroja1997,Huang2015,Sibille2015,Liu2016,Higuchi2016,Shin2018}. \CIG\ is one of these systems and is a prototypical example of case iii)~\cite{Rai2018}. Since the CEF has a very strong influence on the physical properties, especially in case iii), it is very important to determine the CEF scheme of a compound, i.e., the wave functions and the excitation energies of the different CEF levels. A standard approach is, e.g., to fit the anisotropy of the magnetic susceptibility over a wide $T$ range. For the general case, solving the CEF problem and calculating the magnetization implies solving a matrix of size $N \times N$, where $N$ is the degeneracy of the ground state $J$ multiplet: Thus, for Ce$^{3+}$, $N = 6$. However, in some cases in which the Ce site has a higher point symmetry, the problem can be highly simplified. A first simplification is to solve the CEF problem not at a finite magnetic field but at $B = 0$, and then to calculate the susceptibility using perturbation theory up to the second order. First and second order correspond to the Curie and the Van Vleck contributions, respectively. Because Ce$^{3+}$ with an odd number of $f$-electrons is a Kramers system, the size of the relevant matrix is reduced to $3 \times 3$. For some high symmetry cases, only one of the non-diagonal CEF parameters is allowed. As a consequence, only two of the $\left|5/2,m_{z}\right>$ states mix, while the third one remains a pure one. Accordingly, the $3 \times 3$ matrix reduces to only one non-zero (diagonal) element and a $2 \times 2$ block, which can easily be solved analytically. This is, e.g., the case for Ce$^{3+}$ in a tetragonal environment, for which simple analytical solutions are available in the literature~\cite{Fischer1987}. Ce$^{3+}$ in a trigonal environment is also such a case for which a manageable analytic solution can be obtained. However, because in the past the number of trigonal Ce-based systems was very limited, this analytical solution has not been published, yet. In the present paper we provide the general solution for the calculation of the CEF scheme and of the susceptibility of Ce in a trigonal environment. We use this general solution to analyze the anisotropic susceptibility of the two recently reported compounds \CIG~\cite{Rai2018} and \CCA~\cite{Liu2016}. In \CIG\ we find an unusual large CEF splittings, with the second excited CEF level at about 1400\,K. In \CCA\ we show that its highly anisotropic susceptibility can be perfectly reproduced by the CEF scheme elaborated here using our general solution. This demonstrates that this compound is an easy plane system with a comparatively small exchange interaction and not an Ising system with a huge anisotropic exchange, as originally proposed in Ref.~\onlinecite{Liu2016}. We compare the results for \CIG\ and \CCA\ with two further systems with trigonal symmetry, CeAuSn~\cite{Adroja1997,Huang2015} and \CPAG~\cite{Shin2018}, and show that this point symmetry generally results in a strong easy plane anisotropy and quite similar CEF schemes, making the CEF of \CIG\ an exemplary case.
\section{Experimental techniques}
\begin{figure}[t]
	\centering
	\includegraphics[width=\columnwidth]{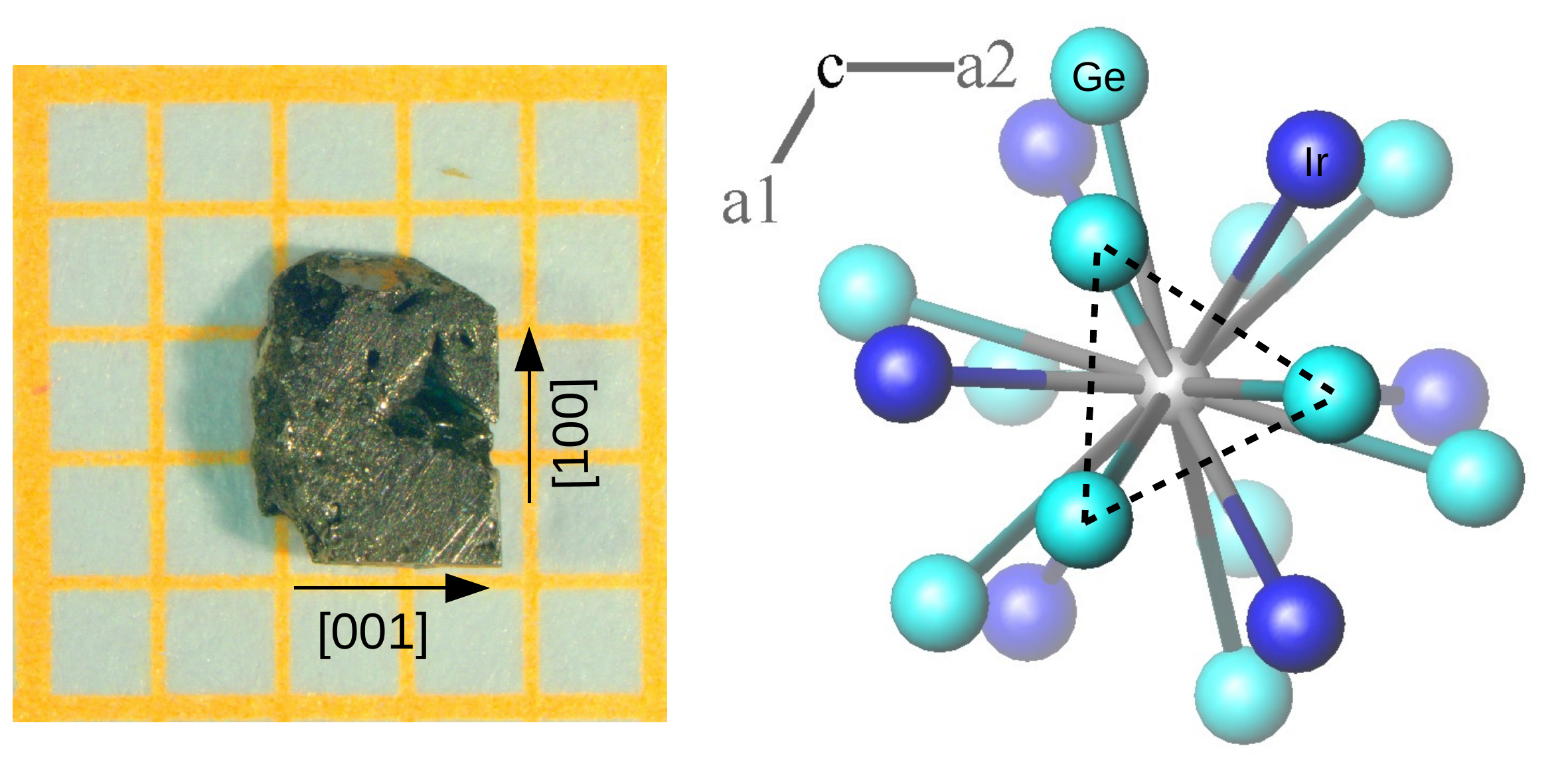}
	\caption{Left: Optical microscope image of a \CIG\ single crystal grown using self flux method~\cite{Rai2018}. Right: The Ce-atom (gray) coordination in \CIG, which emphasizes its trigonal point symmetry. There are six Ir neighbors (distance 3.21~\AA) and twelve Ge neighbors (2$\times$6, distances 3.25~\AA\ and 3.26~\AA).}
	\label{fig1}
\end{figure}

For this work we have used single crystals (see Fig.~\ref{fig1}) that were grown using self-flux technique~\cite{Rai2018}. X-ray diffraction was used for the identification of phase purity of the crystals and Laue method of back scattering reflection was used for the orientation. Using a superconducting quantum interference device (SQUID) and vibrating sample magnetometer (VSM), dc magnetic susceptibility and magnetization were measured between 1.8 and 600\,K and in magnetic fields up to 7\,T. A comprehensive study of transport, thermodynamic and magnetic properties of \CIG\ is presented in Ref.~\cite{Rai2018}. 
\section{Results}
\CIG\ is one compound of a poorly studied RT$_{3}$M$_{7}$ (R: rare earth, T: transition metal and M: XIV group element) family which crystallizes in the rhombohedral $R\bar{3}c$ space group (isostructural to ScRh$_{3}$Si$_{7}$~\cite{Chabot1981}) with lattice parameters $a = 7.89$\,\AA\ and $c = 20.78$\,\AA. The Ce atom has a single site at the $6b$ position with trigonal point symmetry $\bar{3}$ (see Fig.~\ref{fig1}). The structure is centrosymmetric.
		
\begin{figure}[t]
	\centering
	\includegraphics[width=\columnwidth,angle=0]{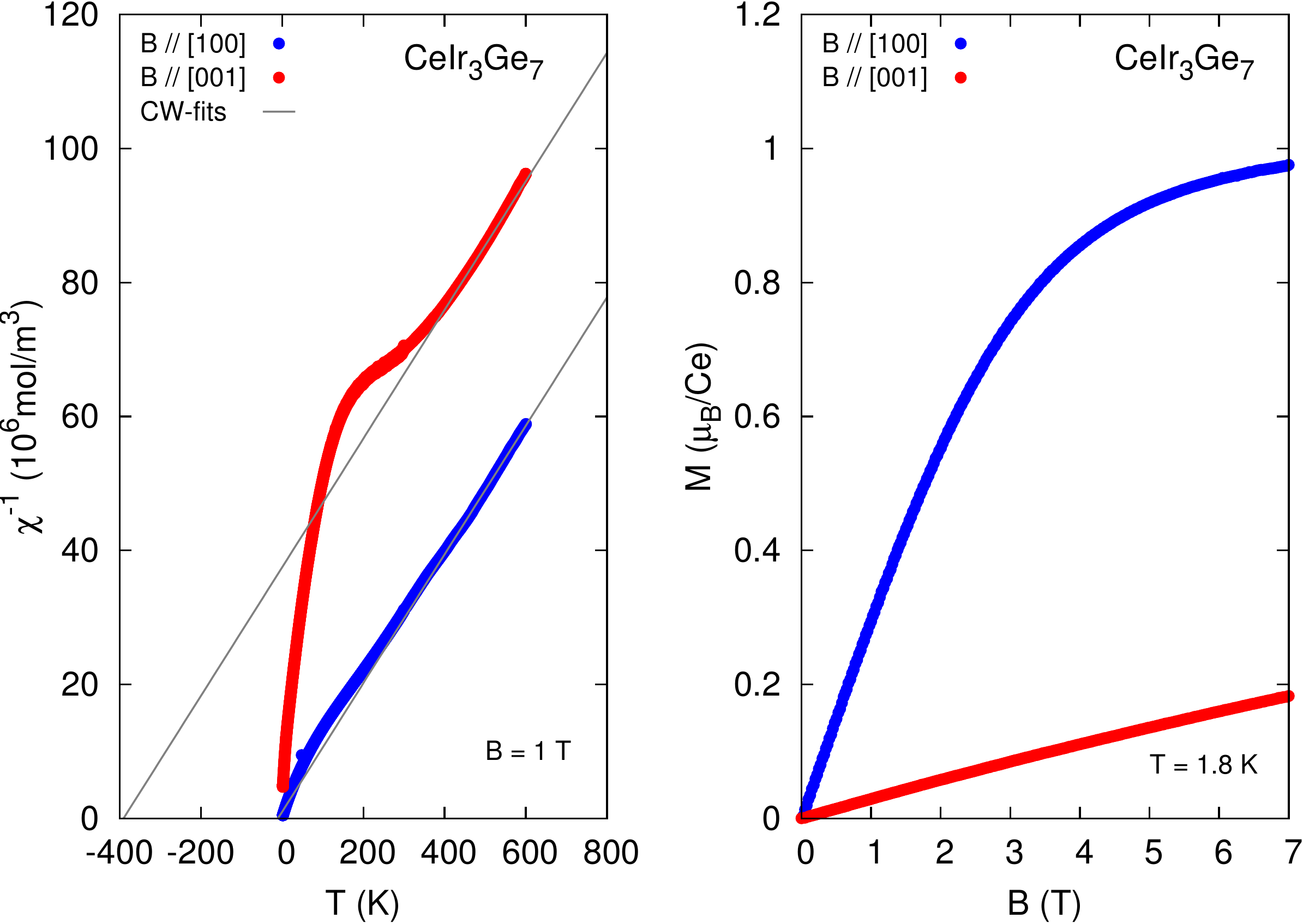}
	\caption{Left: Inverse magnetic susceptibility of \CIG\ measured at $B = 1$\,T applied along the [100] and [001] crystallographic axes. The grey lines are linear fits to Curie-Weiss law at temperatures between 400 and 500\,K. Right: Field dependence of the magnetization measured at 1.8\,K along the [100] and [001] axes.}
	\label{fig2}
\end{figure}

This system is paramagnetic down to the AFM transition temperature of $0.63$\,K and presents a large magnetocrystalline anisotropy~\cite{Rai2018}. This is evidenced by the inverse magnetic susceptibility $\chi^{-1} = B / M$ shown in Fig.~\ref{fig2} (left) between 1.8 and 600\,K. The crystallographic $c$-axis is the magnetic hard axis. Above 400\,K, $\chi^{-1}$ follows a Curie-Weiss (CW) behaviour along both field directions. Fitting the data between 400 and 500\,K, the CW law yields an effective moment \mueff\ = $(2.52 \pm 0.1)$\,\muB, very close to that of the free Ce$^{3+}$ ion of 2.54\muB. We fit only in this temperature range because of the slight upturn of $\chi^{-1}$ for $B \parallel [001]$ above 500\,K, which is emphasized in Fig.~\ref{fig4} (right). This indicates an additional diamagnetic contribution $\chi_{0}$ which is possibly due to the sample holder. This contribution was found to be $\chi_{0} = -2\times10^{-10}$\,m$^{3}$/mol. If subtracted, the high temperature CW fit yields an effective moment of 2.54\,\muB\ as expected for a pure Ce$^{3+}$ ion (cf. Fig.~\ref{fig5}, left). 

Since the point symmetry of the Ce atom is trigonal, the CEF Hamiltonian has just three parameters. Therefore, to solve exactly the CEF scheme it would be enough to fit the temperature dependence of the susceptibility and the field dependence of the magnetization at low temperature along both principal crystallographic axes. In fact, magnetization at 1.8\,K (Fig.~\ref{fig2}, right) suggests a saturation moment of about 1\,\muB\ along [100] and much smaller along [001] for the ground state wave function. Before solving the Hamiltonian we can obtain an estimation of the CEF $B_{2}^{0}$ parameter by using the preliminary CW fits shown in Fig.~\ref{fig2} (left): The paramagnetic Weiss temperatures along both principal crystallographic axes [100] and [001] are $\theta_{W}^{a} = 9.5$\,K and $\theta_{W}^{c} = -361$\,K, respectively. Since the ordering temperatures and thus exchange interactions in Ce-based systems are comparatively weak ($< 10$\,K) because of the tiny de Gennes factor, the anisotropy of $\chi(T)$ at high $T$ is dominated by the effect of the crystalline electric field. The Weiss temperatures can be then expressed on the basis of a high-temperature series expansion as a function of the first CEF parameter $B_{2}^{0}$~\cite{Bowden1971}, which is therefore a measure of the strength of the magnetocrystalline anisotropy:
\begin{equation}\label{eq:B20}
	B_{2}^{0} = \left(\theta_{W}^{a} - \theta_{W}^{c}\right) \frac{10k_{B}}{3(2J-1)(2J+3)}.
\end{equation}
Using the paramagnetic Weiss temperatures, we find that $B_{2}^{0}$ = 3.32\,meV = 38.5\,K. This value is consistent with the large difference between the saturation moments along the [100] and [001] directions observed in magnetization. 

\begin{figure}[t]
	\centering
	\includegraphics[width=\columnwidth,angle=0]{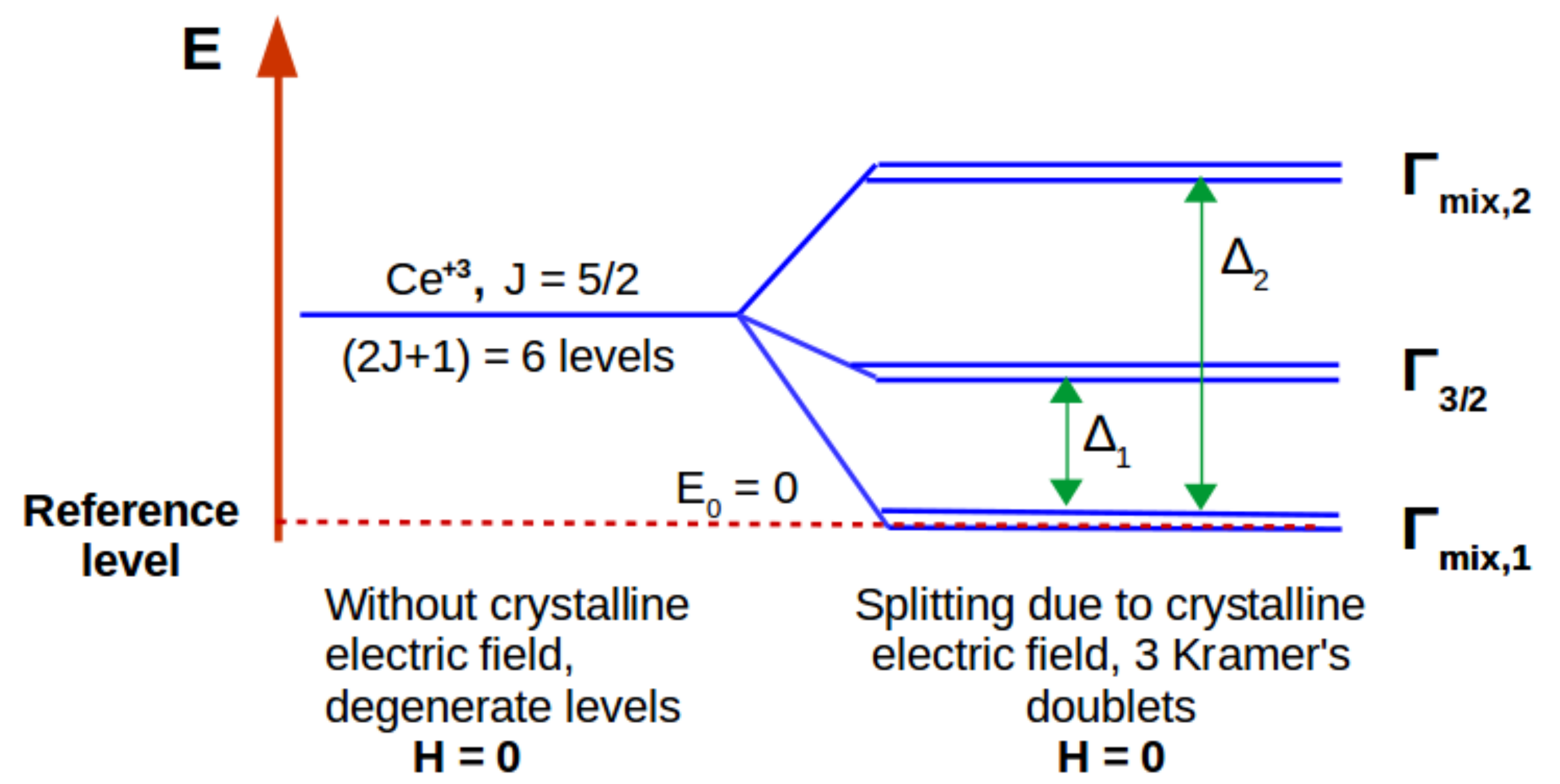}
	\caption{Proposed CEF scheme for \CIG\ with $\Delta_{1}=374$\,K,  and $\Delta_{2}=1398$\,K.}
	\label{fig3}
\end{figure}

\begin{figure}[t]
	\centering
	\includegraphics[width=\columnwidth,angle=0]{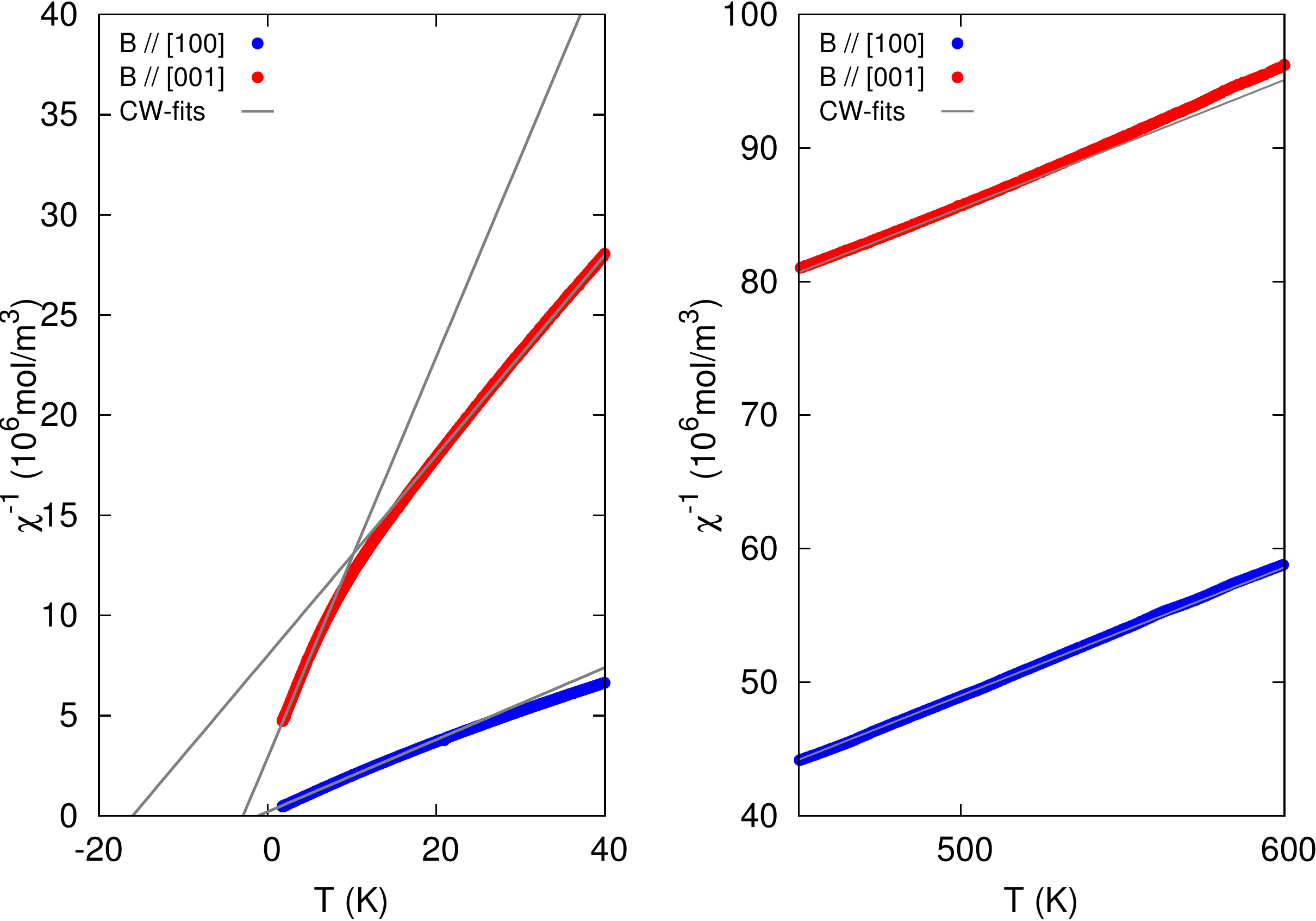}
	\caption{Zoom of the inverse magnetic susceptibility of \CIG\ at low (left) and high (right) temperatures. The grey lines are linear fits to Curie-Weiss law which evidence a kink at about 10\,K (left) possibly due to a small additional paramagnetic contribution and an upturn above 500\,K due to a diamagnetic contribution from the sample holder. }
	\label{fig4}
\end{figure}

\begin{figure}[t]
	\centering
	\includegraphics[width=\columnwidth,angle=0]{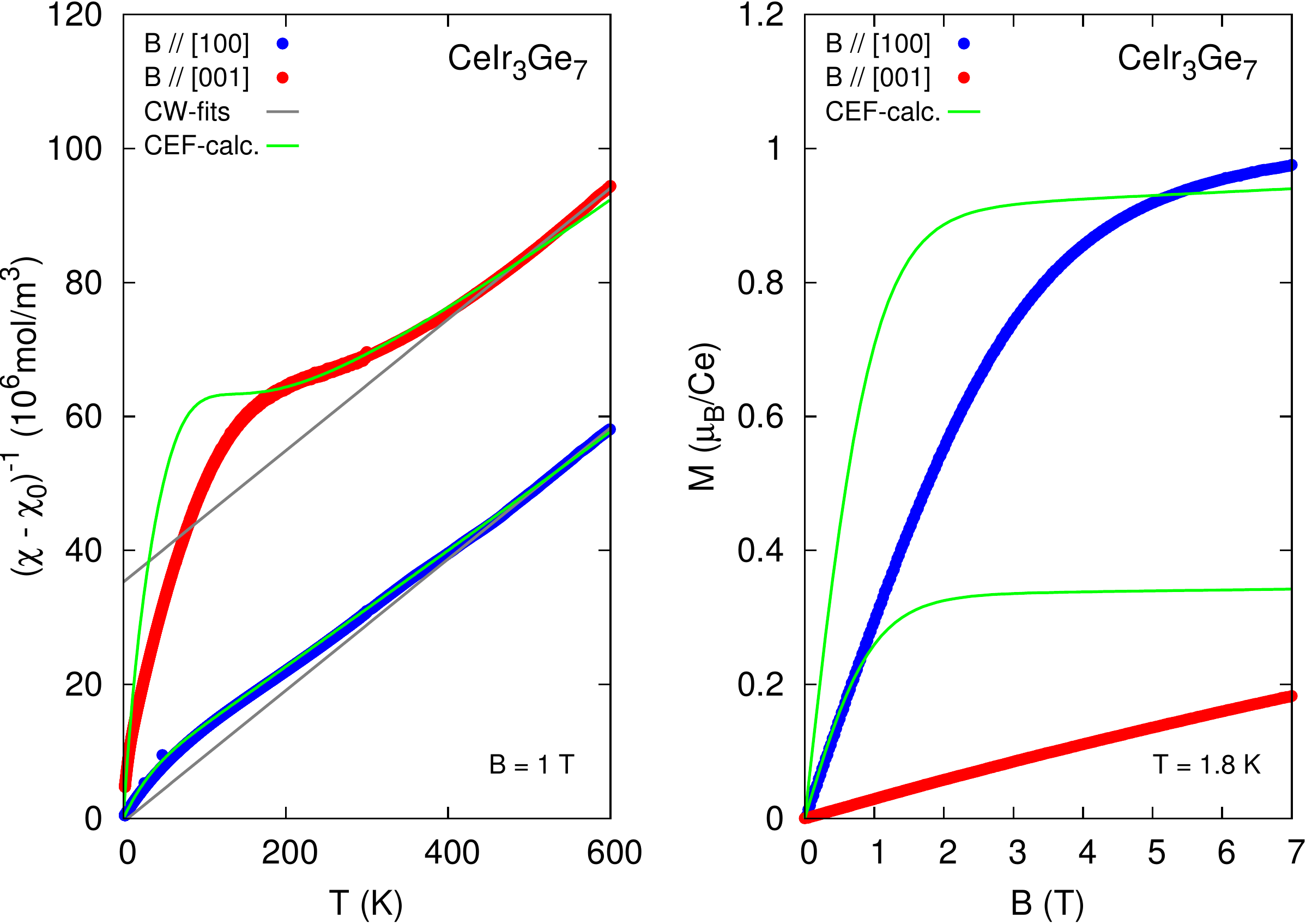}
	\caption{Left: Inverse magnetic susceptibility of \CIG\ measured at $B = 1$\,T after subtraction of the diamagnetic $\chi_{0}$ contribution. The grey lines are linear fits to Curie-Weiss law between 400 and 600\,K which yield an effective moment of 2.54\,\muB, as expected for Ce$^{3+}$. The CW temperatures are $\theta_{W}^{a} = 5$\,K and $\theta_{W}^{c} = -362$\,K which give (from Eq.~\ref{eq:B20}) $B_{2}^{0} = 38.1$\,K. Right: Field dependence of the magnetization measured at 1.8\,K. In both plots, the green lines are the results of the CEF calculations after having fitted $(\chi - \chi_{0})^{-1}$ in the ranges 1.8 - 6\,K and 200 - 600\,K for $B \parallel [001]$ and in the whole measured range for $B \parallel [100]$.}
	\label{fig5}
\end{figure}

We can now use this value as a starting point for evaluating the CEF scheme. For the trigonal point symmetry of the Ce atoms in the crystal, the $(2J + 1)$ six-fold degenerate levels split into three Kramers doublets (see Fig.~\ref{fig3}). The CEF Hamiltonian is given by
\begin{equation*}
	H_{CEF} =  B_{2}^{0}O_{2}^{0} + B_{4}^{0}O_{4}^{0} + B_{4}^{3}O_{4}^{3}
\end{equation*} 
where $B_{m}^{n}$ are CEF parameters and $O_{m}^{n}$ are Steven operators~\cite{Stevens1952,Hutchings1964} which are given by
%
\begin{align*}
O_{2}^{0}\left|J,m\right> = & [3J_{z}^{2} - J(J+1)] \\
O_{4}^{0}\left|J,m\right> = & [35J_{z}^{4} - 30J(J+1)J_{z}^{2} + 25J_{z}^{2} \\
& - 6J(J+1) + 3J^{2}(J+1)^{2}] \\
O_{4}^{3}\left|J,m\right> = & \frac{1}{4}[J_{z}(J_{+}^{3} + J_{-}^{3}) + (J_{+}^{3} + J_{-}^{3})J_{z}]
\end{align*}
%
with operators
\begin{align*}
& J_{z}\left|j,m\right> = m\left|j,m\right> \\
& J_{x}\left|j,m\right> = \frac{1}{2}\left(J_{+} + J_{-}\right)\left|j,m\right> \\
& J_{\pm}\left|j,m\right> = \left[j(j+1)-m(m\pm 1)\right]^{1/2}\left|j,m\pm 1\right>.
\end{align*}
For the following calculations we define the $[001]$ crystallographic direction as the quantisation axis $z$ and the [100] direction as the $x$ axis. Adding the Zeeman term, the global Hamiltonian is
\begin{equation}\label{eq:Ham}
H_{CEF} =  B_{2}^{0}O_{2}^{0} + B_{4}^{0}O_{4}^{0} + B_{4}^{3}O_{4}^{3} + g_{J}J\muB B
\end{equation} 
with $g_{J} = 6/7$ the Land\'e $g$-factor for Ce and $\mu_{B} = 9.274\times10^{-24}$\,J/T the Bohr magneton.

The CEF Hamiltonian matrix can be calculated by noting down all the non-zero matrix elements~\cite{Hutchings1964}. After rearranging the states we obtain
\begin{center}
	\begin{tabular}{c|ccc}
	\large
	$H_{CEF}$ & $\left|\pm\frac{5}{2}\right>$ & $\left|\mp\frac{1}{2}\right>$ & $\left|\pm\frac{3}{2}\right>$ \\ [5pt] \hline \\
	$\left<\pm\frac{5}{2}\right|$ & $A$ & $D$ & 0 \\[20pt]
	$\left<\mp\frac{1}{2}\right|$ & $D$ & $B$ & 0 \\[20pt]
	$\left<\pm\frac{3}{2}\right|$ & 0 & 0 & $C$ \\[8pt]
	\end{tabular}
\end{center}
where we have defined
\begin{align*}
& A = (10B_{2}^{0} + 60B_{4}^{0}) \\
& B = (-8B_{2}^{0} + 120B_{4}^{0}) \\
& C = (-2B_{2}^{0} - 180B_{4}^{0}) \\
& D = \pm(3\sqrt{10}B_{4}^{3}).
\end{align*}
The solutions are three doublets of the form
\begin{align*}
& \left|\Gamma_{mix,1}\right> = \cos\alpha\left|\pm 5/2\right> + \sin\alpha\left|\mp 1/2\right> \\
& \left|\Gamma_{mix,2}\right> = \sin\alpha\left|\pm 5/2\right> - \cos\alpha\left|\mp 1/2\right> \\
& \left|\Gamma_{3/2}\right> = \left|\pm 3/2\right>
\end{align*}
with eigenvalues
\begin{align*}
\epsilon_{mix} = & ~\frac{1}{2}\left[(A + B)\pm\sqrt{(A-B)^{2} + 4D^{2}}\right] = \\ & ~\Gamma_{mix} \pm \sqrt{S_{mix}^{2} + D^{2}} \\
\epsilon_{3/2} = & ~\left<\Gamma_{3/2}|H_{CEF}|\Gamma_{3/2}\right>  = C
\end{align*}
where we define
\begin{align*}
& \Gamma_{mix} = (A + B)/2 = (B_{2}^{0} + 90B_{4}^{0}) \\
& S_{mix} = (A - B)/2 = (9B_{2}^{0} - 30B_{4}^{0}) \\
& \tan\alpha = (-S_{mix}/{D}) \pm \sqrt{(S_{mix}/D)^{2} + 1}.
\end{align*}
Knowledge of $B_{2}^{0}$, $B_{4}^{0}$, $B_{4}^{3}$ allows to simply calculate all other quantities. Our fit to the susceptibility and magnetization data (see below) suggests a CEF level scheme for \CIG\ as the one shown in Fig.~\ref{fig3} with the $\left|\Gamma_{mix,1}\right>$ as ground state and $\left|\Gamma_{3/2}\right>$ as first excited state.

The theoretical expression for the magnetic susceptibility $\chi$ with few approximations at different energy levels by Van Vleck is given by
\begin{widetext}
\begin{equation*}
\chi = \frac{2N_{A}g_{J}^{2}\mu_{B}^{2}\mu_{0}}{Z}\left[
\sum\limits_{n}\beta |\left< J_{i,n} \right> | ^{2}e^{-\beta E_{n}}
+ 2\sum\limits_{m \neq n} |\left< m|J_{i,n}|n \right>|^{2} \left(\frac{e^{-\beta E_{m}} - e^{-\beta E_{n}}}{E_{n} - E_{m}} \right)\right]
\end{equation*}
\end{widetext}
with $\beta = 1/k_{B}T$, $Z = 2\sum\limits_{n}e^{-\beta E_{n}}$, $i=x,z$ and $n,m = 0, 1, 2$.
Here $z$ is the quantization axis, $N_{A}= 6.023 \times 10^{23}$\,/mol the Avogadro number, $k_{B} = 1.38 \times 10^{-23}$\,J/K the Boltzmann constant and $\mu_{0} = 4\pi \times 10^{-7}$\,N/A$^{2}$. The first term is the Curie contribution to the paramagnetic susceptibility and the second term is the Van Vleck susceptibility. For instance, the Curie contribution to the paramagnetic susceptibility for the proposed scheme along both applied field directions can be written as
\begin{widetext}
\begin{align*}
& \chi_{B\parallel z}^{para} = \frac{2\beta N_{A}g_{J}^{2}\mu_{B}^{2}\mu_{0}}{Z} \left[\left<J_{z0}\right>^{2} + \left<J_{z1}\right>^{2}e^{-\beta E_{1}} + \left<J_{z2}\right>^{2}e^{-\beta E_{2}}\right] \\
& \chi_{B\perp z}^{para} = \frac{2\beta N_{A}g_{J}^{2}\mu_{B}^{2}\mu_{0}}{Z} \left[\left<J_{x0}\right>^{2} + \left<J_{x1}\right>^{2}e^{-\beta E_{1}} + \left<J_{x2}\right>^{2}e^{-\beta E_{2}}\right] \\
& Z = 2\left(1 + e^{-\beta E_{1}} + e^{-\beta E_{2}}\right).
\end{align*}
\end{widetext}
After substituting the corresponding expectation values (see Appendix), the final paramagnetic and Van Vleck susceptibilities are given by 
\begin{widetext}
\begin{align*}
& \chi_{B\parallel z}^{para} = \frac{\beta N_{A}g_{J}^{2}\mu_{B}^{2}\mu_{0}}{2Z}\left[\left(5\cos^{2}\alpha - \sin^{2}\alpha\right)^{2} + 9e^{-\beta E_{1}} + \left(5\sin^{2}\alpha - \cos^{2}\alpha\right)^{2}e^{-\beta E_{2}}\right]\\
& \chi_{B\perp z}^{para} = \frac{\beta N_{A}g_{J}^{2}\mu_{B}^{2}\mu_{0}}{2Z}\left[\sin^{4}\alpha + 9\cos^{4}\alpha e^{-\beta E_{2}}\right]\\
& \chi_{B\parallel z}^{VV} = \frac{N_{A}g_{J}^{2}\mu_{B}^{2}\mu_{0}}{k_{B}Z} \left[{36\sin^{2}\alpha\cos^{2}\alpha\frac{1-e^{-\beta E_{2}}}{E_{2}}}\right]\\
& \chi_{B\perp z}^{VV} = \frac{N_{A}g_{J}^{2}\mu_{B}^{2}\mu_{0}}{k_{B}Z}\left[ \left(5\cos^{2}\alpha+8\sin^{2}\alpha\right)\frac{1-e^{-\beta E_{1}}}{E_{1}}\right.\left.+9\sin^{2}\alpha\cos^{2}\alpha\frac{1-e^{-\beta E_{2}}}{E_{2}}+\left(5\sin^{2}\alpha+8\cos^{2}\alpha\right)\frac{e^{-\beta E_{1}} - e^{-\beta E_{2}}}{E_{2} - E_{1}}\right].
\end{align*}
\end{widetext}
The total susceptibilities are then given by:
\begin{equation}\label{eq:chi}
	\chi_{B\parallel z}^{total} = \chi_{B\parallel z}^{para} + 	\chi_{B\parallel z}^{VV} \qquad
	\chi_{B\perp z}^{total} = \chi_{B\perp z}^{para} +	\chi_{B\perp z}^{VV}.
\end{equation}
These equations were used to fit the temperature dependence of the susceptibility of \CIG\ measured at 1\,T (see Fig.~\ref{fig5}).

Before fitting the data, it is useful to take a closer look at the inverse magnetic susceptibility $\chi^{-1}$ measured with $B \parallel [001]$ at low temperatures. This is because $\chi^{-1}$ along this direction is strongly temperature dependent and a fit to the data can already deliver correct CEF parameters. $\chi^{-1}$ vs. $T$ below 40\,K is shown in the left panel of Fig.~\ref{fig4}. $\chi^{-1}$ increases linearly with $T$ between 1.8 and 6\,K with a slope of $1\times10^{6}$\,mol/m$^{3}$K indicated by a grey line. This slope yields a CW effective moment of 0.8\,\muB\ and saturation moment of 0.46\,\muB\ which would be in agreement with the saturation magnetization expected from the field dependence of the magnetization measured at 1.8\,K (cf. Fig.~\ref{fig1}). However, at about 10\,K we notice a significant change in slope into another linear increase with $0.5\times10^{6}$\,mol/m$^{3}$K which extends to temperatures above 40\,K. This would yield a saturation moment of 0.65\,\muB\ which seems to be much too high when compared with the measured magnetization at 1.8\,K and with the value of $B_{2}^{0}$ extracted from the Weiss temperatures.
A fit of  $\chi^{-1}(T)$ for $B \parallel [100]$ at low $T$ gives a saturation moment of 1.1\,\muB\ which also agrees with the magnetization measured at 1.8\,K. This implies that there is an additional paramagnetic contribution, possibly from a secondary phase, which affects the susceptibility along the hard axis at temperatures above 10\,K. Above about 150\,K its magnitude becomes negligible. For this reason, we fit our data along the [001] direction with equations~\ref{eq:chi} in the temperature ranges 1.8 - 6\,K and 200 - 600\,K. This procedure has been found to be correct, since a fitting of the susceptibility along the [001] direction between 10 and 150\,K (with the intention to fit well the hump in $\chi^{-1}(T)$ in Fig.~\ref{fig2}), would give a value of $B_{2}^{0} < 30$\,K which can not reproduce either the susceptibility results at high $T$ or the low-$T$ saturation magnetization. We also considered a possible misalignment of the sample and tried a fit with different weights for both crystallographic directions, but we were never been able to reproduce the data correctly, since this change of slope at 10\,K is too pronounced to be reproduced by a misalignment.

The fitted functions together with the experimental data are shown in Fig.~\ref{fig5} for both field directions. In our calculation the quantization $z$-axis is the experimental crystalline [001] direction. We could perfectly fit $\chi^{-1}(T)$ with $B \parallel [100]$ in the whole temperature range, the high-temperature part of $\chi^{-1}(T)$ with $B \parallel [001]$ as well as its low-$T$ part with a single set of CEF parameters. These parameters are listed in Tab.~\ref{tab}. The $B_{2}^{0} = 34.4$\,K is a bit smaller that evaluated from Eq.~\ref{eq:B20} using the CW temperatures from fits shown in Fig.~\ref{fig5}, which is 38.1\,K. With the same set of parameters and the Zeeman energy in the Hamiltonian (Eq.~\ref{eq:Ham}) we calculated the magnetization at 1.8\,K which is shown in the right panel of Fig.~\ref{fig5}. The evolution of the magnetization in field does not correspond well to what has been measured, because we have not considered any exchange term in the Hamiltonian: Comparing the initial slope $M/B$ of the experimental data with that of the calculation we can provide an estimation of the exchange interaction which resulted to be 2.4\,K, i.e., comparable with the Weiss temperature extracted from the Curie-Weiss fit of the average inverse susceptibility~\cite{Rai2018}. The saturation moments agree well with those measured by experiments. The small difference is due other paramagnetic contributions as, e.g., from $5d$ electrons.
\begin{table}[ht!]
	\centering
	\caption{Crystalline electric field parameters for \CIG\ and \CCA. We consider the reference level $E_{0} = 0$ and $E_{1} = \Delta_{1}$, $E_{2} = \Delta_{2}$.}
	\label{tab}
	\begin{ruledtabular}
		\CIG\\
	\begin{tabular}{lll}
		CEF parameters & Energies & Mixing angle\\
		\hline \\
		$B_{2}^{0} = 34.4$\,K & $\Delta_{1} = 374$\,K & $\alpha = -57\degree$ \\
		$B_{4}^{0} = 0.82$\,K & $\Delta_{2} = 1398$\,K \\
		$B_{4}^{3} = 67.3$\,K \\
	\end{tabular}
	\end{ruledtabular}
		\begin{ruledtabular}
			\\
			\CCA\\
			\begin{tabular}{lll}
				CEF parameters & Energies & Mixing angle\\
				\hline \\
				$B_{2}^{0} = 11.6$\,K & $\Delta_{1} = 241$\,K & $\alpha = -73.8\degree$ \\
				$B_{4}^{0} = -0.5$\,K & $\Delta_{2} = 282$\,K \\
				$B_{4}^{3} = 8.0$\,K \\
			\end{tabular}
		\end{ruledtabular}
\end{table}
In fact, the CEF parameters leave $\left|\Gamma_{mix,1}\right>$ as the ground state wave function with a mixing angle $\alpha = -57\degree$ and energy splittings between the ground state and the first and second excited states of $\Delta_{1} = 374$\,K\,$\approx$\,32\,meV and $\Delta_{2} = 1398$\,K\,$\approx$\,120\,meV. These energy splittings are extremely large when compared with other Ce-based intermetallics which commonly have splittings between 10 and 60\,meV~\cite{Christianson2004,Pikul2010,Willers2012}. 

With  this ground state function, $\left|\Gamma_{mix,1}\right> = 0.54\left|\pm 5/2\right> - 0.84\left|\mp 1/2\right>$, we can easily calculate the saturation magnetization with field along both crystallographic directions. Considering the Zeeman term $H_{x,z} = g_{J}\mu_{B}\left<J_{x,z}\right>B_{z}$, we have
\begin{align*}
& \left<\Gamma_{mix,1}|J_{z}|\Gamma_{mix,1}\right>g_{J}\mu_{B}B = 0.39g_{J}\mu_{B}B \\ & \left<\Gamma_{mix,1}|J_{x}|\Gamma_{mix,1}\right>g_{J}\mu_{B}B = 1.06g_{J}\mu_{B}B
\end{align*}
The saturation magnetizations parallel and perpendicular to the $z$-axis are
\begin{align*}
& M_{B\parallel z} =  0.39\left(6/7\right)\mu_{B} = 0.33\mu_{B} \\
& M_{B\perp z} = 1.06\left(6/7\right)\mu_{B} = 0.91\mu_{B}.
\end{align*}
\begin{figure}[t]
	\centering
	\includegraphics[width=\columnwidth,angle=0]{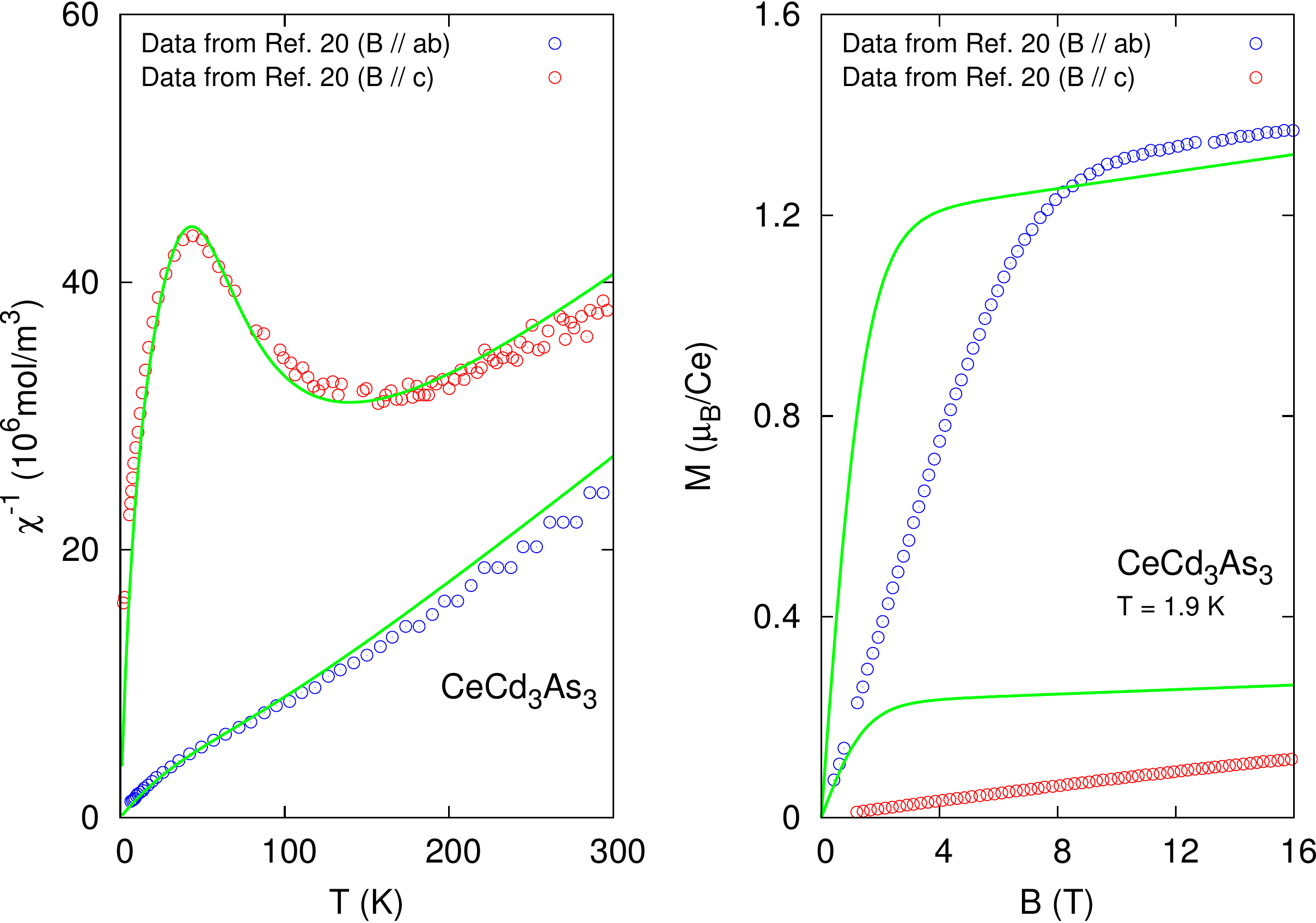}
	\caption{Left: Inverse magnetic susceptibility of \CCA. Right: Field dependence of the magnetization measured at 1.9\,K. The data were taken from Ref.~\cite{Liu2016}. The green lines are our CEF calculations with parameters given in the text and in Tab.~\ref{tab}.}
	\label{fig6}
\end{figure}

To show the validity and the general character of our calculation we apply the same procedure to the recently discovered system \CCA~\cite{Liu2016}. The huge anisotropy of the susceptibility of this compound, as well as the weak $T$ dependence of its $c$-axis susceptibility led the authors of Ref.~\onlinecite{Liu2016} to propose that \CCA\ is a very strong Ising type system with a huge anisotropy of the exchange interaction; the exchange along $c$ being orders of magnitude larger than in the basal plane. However, these authors did not try to analyse their data using a CEF model. Susceptibility and magnetization data taken from Ref.~\cite{Liu2016} are plotted in Fig.~\ref{fig6}. We fit these data with our model and found a very good agreement: The CEF parameters are listed in Tab.~\ref{tab} and leave $\left|\Gamma_{mix,1}\right>$ as the ground state wave function with a mixing angle $\alpha = -73.8\degree$ and energy splittings between the ground state and the first and second excited states of $\Delta_{1} = 241$\,K\,$\approx$\,20.8\,meV and $\Delta_{2} = 282$\,K\,$\approx$\,24.3\,meV. The saturation magnetizations parallel and perpendicular to the $c$-axis are $M_{B\parallel c} =  0.23\mu_{B}$ and $M_{B\perp c} = 1.19\mu_{B}$. Thus, our analysis shows that the strongly anisotropic susceptibility of \CCA\ and its peculiar $T$ dependence of the $c$-axis susceptibility can be fully accounted for by the CEF, with a CEF scheme quite similar to that of \CIG, except for a smaller overall splitting. Therefore, instead of being a strongly Ising type system, \CCA\ is an easy plane XY system with a standard strength of the exchange interaction along both directions. This case demonstrate the importance of doing a CEF analysis before discussing the properties of a rare-earth-based magnetic system, and the value of a general analytical solution to the CEF problem.

In the course of our study we noticed two further systems with Ce$^{3+}$ in a trigonal (local) environment, CeAuSn~\cite{Adroja1997,Huang2015} and \CPAG~\cite{Shin2018}. Both show an anisotropy very similar to that of \CIG, with a large easy plane susceptibility and a small $c$-axis CEF ground state moment. The CEF of CeAuSn has been well-analyzed in two successive papers~\cite{Adroja1997,Huang2015} leading to a convergent solution quite similar to that of \CIG, except for a much smaller overall splitting. In contrast, for \CPAG\ no CEF analysis was performed, but the similarity of its susceptibility data to those of \CCA\ implies a very similar CEF scheme, too. Thus, all the trigonal Ce-based systems investigated recently bear a very similar CEF scheme with a very pronounced easy plane anisotropy and a small $c$-axis CEF ground state moment. This origins from a large positive $B_{2}^{0}$ coefficient, but also from a large mixing coefficient $B_{4}^{3}$, which is of the same order or even larger than $B_{2}^{0}$, resulting in a large mixing between the $\left|\mp 1/2\right>$ and the $\left|\pm 5/2\right>$ states. This is a fundamental difference to purely hexagonal systems with a sixfold point symmetry, where the mixing term is absent, resulting in pure $\left|\pm 1/2\right>$, $\left|\pm 3/2\right>$ and $\left|\pm 5/2\right>$ CEF doublets.
\section{Discussion and Conclusion}
A comprehensive analysis of the CEF scheme of Ce$^{3+}$ in the trigonal point symmetry has been presented. We provided a general analytic solution which can be used to solve the CEF problem and to calculate the anisotropic magnetic susceptibility for Ce in a trigonal surrounding. We have successfully used this solution to analyze the susceptibility of the new compound \CIG\ and to determine its CEF scheme. This analysis indicates that the ground state doublet in this compound is composed by a large mixing of the $\left|\pm 5/2\right>$ and the $\left|\mp 1/2\right>$ $m_{z}$ states, and the first and second excited states are at 374\,K and 1398\,K, respectively. The latter value is exceptionally large compared to the typical values of $200-600$\,K observed in intermetallic Ce-based compounds. Further on, we used the same analytical solution to analyze the anisotropic susceptibility of \CCA. We showed that the anisotropic susceptibility of this compound can be fully accounted for by the CEF, providing a much simpler and standard explanation for its peculiar susceptibility than that originally proposed. We found that two further compounds, CeAuSn and \CPAG, presents a very similar anisotropy and accordingly a very similar CEF scheme to that of \CIG. This indicates that the CEF of \CIG\ is an exemplary case for Ce-based systems with trigonal symmetry. This systematic study also shows that in intermetallic Ce-based systems a trigonal environment usually results not only in a large $B_{2}^{0}$ CEF parameter, but also to a large $B_{4}^{3}$ mixing CEF parameter, both leading to a pronounced easy plane behavior.

Our analysis indicates an unusual large overall CEF splitting in \CIG. This huge splittings might be connected with the presence of 5$d$ ligands, i.e., nearest-neighbor iridium atoms (see Fig.~\ref{fig1}). To check this idea, we have performed band structure calculations using the full-potential local-orbital FPLO code~\cite{Koepernik1999}. For the exchange and correlation potential, the local density approximation~\cite{Perdew1992} was applied. The calculations were carried out scalar relativistically on a well converged $k$-mesh (20$\times$20$\times$20). The influence of the spin-orbit coupling to the valence states is rather small. We have used the room-temperature data of Ref.~\cite{Rai2018} for the lattice parameters and treated the cerium $4f$-states as core states (open core approximation).

\begin{figure}[t]
	\centering
	\includegraphics[width=\columnwidth,angle=0]{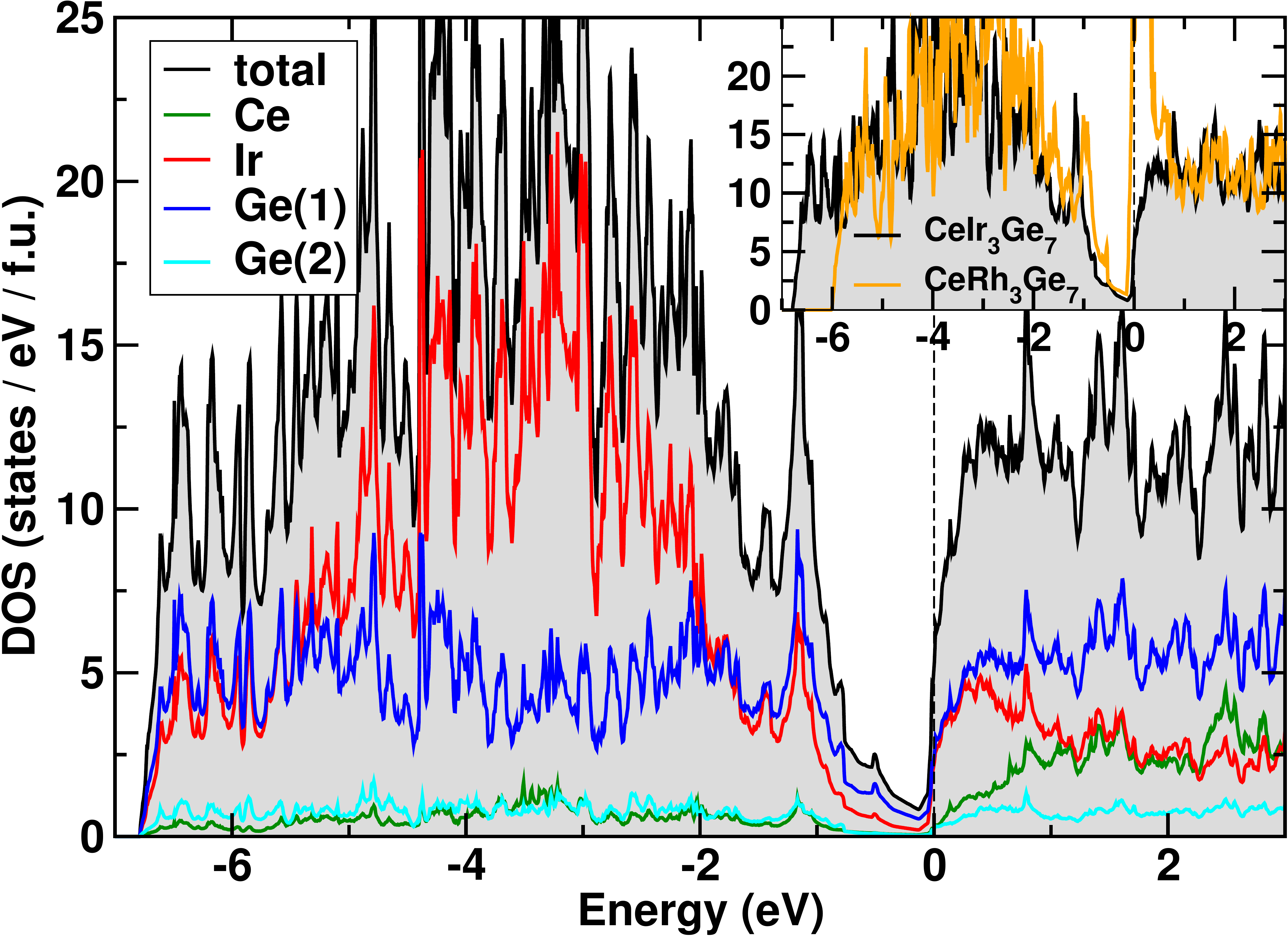}
	\caption{Calculated total and partial density of states (DOS) of \CIG. Both Ir (red) and Ge(1) (blue) contribute almost equally to the DOS at the Fermi level. The inset shows the comparison to the fictitious isostructural Rh system (orange).}
	\label{fig7}
\end{figure}

The calculated density of states (DOS) is shown in Fig.~\ref{fig7}. The valence band is essentially formed by strongly hybridized  Ir (mostly Ir $5d$) and Ge (mostly Ge $4p$) states. A comparison between calculations of \CIG\ and the fictitious \CRG\ (we used the same lattice parameters and Wyckoff positions due to the very small difference in atomic size between Ir and Rh) shows that for the Ir-based system the band width, which is a measure of the hybridization, is significantly larger than in the Rh-based compound, indicating that the $5d$ ligands create a substantially larger crystalline field.
However, the precise estimation of the CEF parameters with density functional theory calculations is very complex and, most importantly, it depends strongly on the hybridization parameter $\Delta$ which is not known for \CIG. The stronger $\Delta$ the larger the crystalline field~\cite{Huesges2018}. A more quantitative description why this field is so large in \CIG\ can not be answered here and is beyond the scope of this paper.
\section{Acknowledgments}
We are indebted to M. O. Ajeesh, R. Cardoso, D.-J. Jang and  J. Sereni for useful discussions, and D. A. Sokolov for having oriented the crystal. Work at Rice University was supported by the Gordon and Betty Moore Foundation EPiQS Initiative through grant GBMF4417.  BKR acknowledges partial support  by a QuantEmX grant from ICAM and the Gordon and Betty Moore Foundation through Grant GBMF5305. EM acknowledges travel support from the Alexander von Humboldt Foundation through the Fellowship for Experienced Researchers.
\appendix*
\section{APPENDIX}
\section{List of matrix elements}
We list here all matrix elements for the calculation of the susceptibility.
\subsection{Matrix elements for the paramagnetic susceptibility}
For the field parallel to $z$-axis:
\begin{align*}
& \left<J_{z0}\right>: \\
& \left<\Gamma_{mix,1}(a)|J_{z}|\Gamma_{mix,1}(a)\right> = \frac{1}{2}\left[5\cos^{2}\alpha - \sin^{2}\alpha\right] \\
& \left<\Gamma_{mix,1}(b)|J_{z}|\Gamma_{mix,1}(b)\right> = \frac{1}{2}\left[\sin^{2}\alpha - 5\cos^{2}\alpha\right]
\end{align*}
\begin{align*}
& \left<J_{z1}\right>: \\
& \left<\Gamma_{3/2}(a)|J_{z}|\Gamma_{3/2}(a)\right> = +\frac{3}{2} \\
& \left<\Gamma_{3/2}(b)|J_{z}|\Gamma_{3/2}(b)\right> = -\frac{3}{2}
\end{align*}
\begin{align*}
& \left<J_{z2}\right>: \\
& \left<\Gamma_{mix,2}(a)|J_{z}|\Gamma_{mix,2}(a)\right> = \frac{1}{2}\left[5\sin^{2}\alpha - \cos^{2}\alpha\right] \\
& \left<\Gamma_{mix,2}(b)|J_{z}|\Gamma_{mix,2}(b)\right> = \frac{1}{2}\left[\cos^{2}\alpha - 5\sin^{2}\alpha\right]
\end{align*}
All other matrix elements are zero:
\begin{align*}
& \left<\Gamma_{mix,1}(a)|J_{z}|\Gamma_{mix,1}(b)\right> = \left<\Gamma_{mix,1}(b)|J_{z}|\Gamma_{mix,1}(a)\right> = 0 \\
& \left<\Gamma_{mix,2}(a)|J_{z}|\Gamma_{mix,2}(b)\right> = \left<\Gamma_{mix,2}(b)|J_{z}|\Gamma_{mix,2}(a)\right> = 0 \\
& \left<\Gamma_{3/2}(a)|J_{z}|\Gamma_{3/2}(b)\right> = \left<\Gamma_{3/2}(b)|J_{z}|\Gamma_{3/2}(a)\right> = 0.
\end{align*}
For field perpendicular to the $z$ axis: $\left<J_{x0}\right>, \left<J_{x1}\right>, \left<J_{x2}\right>$; Eigenfunctions of these operators are linear combinations of $\left|\Gamma_{mix,1}(a)\right>$ and $\left|\Gamma_{mix,1}(b)\right>$, $\left|\Gamma_{3/2}(a)\right>$ and $\left|\Gamma_{3/2}(b)\right>$, $\left|\Gamma_{mix,2}(a)\right>$ and $\left|\Gamma_{mix,2}(b)\right>$ and vice versa.
\begin{align*}
& \left<J_{x0}\right>: \\
& \left|\Gamma_{mix,1}(c)\right> = \frac{1}{\sqrt{2}}\left[\left|\Gamma_{mix,1}(a)\right> + \left|\Gamma_{mix,1}(b)\right>\right] \\
& \left|\Gamma_{mix,1}(d)\right> = \frac{1}{\sqrt{2}}\left[\left|\Gamma_{mix,1}(a)\right> - \left|\Gamma_{mix,1}(b)\right>\right] \\
& \left<\Gamma_{mix,1}(c)\right|J_{x}\left|\Gamma_{mix,1}(c)\right> = +\frac{3}{2}\sin^{2}\alpha \\
& \left<\Gamma_{mix,1}(d)\right|J_{x}\left|\Gamma_{mix,1}(d)\right> =
-\frac{3}{2}\sin^{2}\alpha
\end{align*}
\begin{align*}
& \left<J_{x1}\right>: \\
& \left|\Gamma_{3/2}(c)\right> = \frac{1}{\sqrt{2}}\left[\left|\Gamma_{3/2}(a)\right> + \left|\Gamma_{3/2}(b)\right>\right] \\
& \left|\Gamma_{3/2}(d)\right> = \frac{1}{\sqrt{2}}\left[\left|\Gamma_{3/2}(a)\right> - \left|\Gamma_{3/2}(b)\right>\right] \\
& \left<\Gamma_{3/2}(c)\right|J_{x}\left|\Gamma_{3/2}(c)\right> = 0 \\
& \left<\Gamma_{3/2}(d)\right|J_{x}\left|\Gamma_{3/2}(d)\right> = 0
\end{align*}
\begin{align*}
& \left<J_{x2}\right>: \\
& \left|\Gamma_{mix,2}(c)\right> = \frac{1}{\sqrt{2}}\left[\left|\Gamma_{mix,2}(a)\right> + \left|\Gamma_{mix,2}(b)\right>\right] \\
& \left|\Gamma_{mix,2}(d)\right> = \frac{1}{\sqrt{2}}\left[\left|\Gamma_{mix,2}(a)\right> - \left|\Gamma_{mix,2}(b)\right>\right] \\
& \left<\Gamma_{mix,2}(c)\right|J_{x}\left|\Gamma_{mix,2}(c)\right> = +\frac{3}{2}\cos^{2}\alpha \\
& \left<\Gamma_{mix,2}(d)\right|J_{x}\left|\Gamma_{mix,2}(d)\right> = -\frac{3}{2}\cos^{2}\alpha.
\end{align*}
\subsection{Matrix elements for the Van Vleck susceptibility}		
Here, we assumed the reference level, $E_{0} = 0, E_{1} = \Delta_{1}, E_{2} = \Delta_{2}$. Also we notice that
\begin{align*}
& \left<\Gamma_{mix,1}|J_{\alpha}|\Gamma_{3/2}\right>^{2} = \left<\Gamma_{3/2}|J_{\alpha}|\Gamma_{mix,1}\right>^{2} \\
& \left<\Gamma_{mix,1}|J_{\alpha}|\Gamma_{mix,2}\right>^{2} = \left<\Gamma_{mix,2}|J_{\alpha}|\Gamma_{mix,1}\right>^{2} \\
& \left<\Gamma_{3/2}|J_{\alpha}|\Gamma_{mix,2}\right>^{2} = \left<\Gamma_{mix,2}|J_{\alpha}|\Gamma_{3/2}\right>^{2}.
\end{align*}
Each of these elements has 4 different combinations of mixed states (a) and (b), evaluated each of them with the operators along both field directions, parallel to the $z$ axis and perpendicular to it. For field parallel to the $z$-axis ($J_{z}$):
\begin{align*}
& \left<\Gamma_{mix,1}|J_{z}|\Gamma_{3/2}\right>: \\
& \left<\Gamma_{mix,1}(a)|J_{z}|\Gamma_{3/2}(a)\right> = \left<\Gamma_{3/2}(a)|J_{z}|\Gamma_{mix,1}(a)\right> = 0\\
& \left<\Gamma_{mix,1}(a)|J_{z}|\Gamma_{3/2}(b)\right> = \left<\Gamma_{3/2}(b)|J_{z}|\Gamma_{mix,1}(a)\right> = 0 \\
& \left<\Gamma_{mix,1}(b)|J_{z}|\Gamma_{3/2}(a)\right> = \left<\Gamma_{3/2}(a)|J_{z}|\Gamma_{mix,1}(b)\right> = 0 \\
& \left<\Gamma_{mix,1}(b)|J_{z}|\Gamma_{3/2}(b)\right> = \left<\Gamma_{3/2}(b)|J_{z}|\Gamma_{mix,1}(b)\right> = 0
\end{align*}
\begin{align*}
& \left<\Gamma_{mix,1}|J_{z}|\Gamma_{mix,2}\right>: \\
& \left<\Gamma_{mix,2}(a)|J_{z}|\Gamma_{mix,1}(a)\right> = +3 \sin\alpha \cos\alpha \\
& \left<\Gamma_{mix,1}(a)|J_{z}|\Gamma_{mix,2}(a)\right> = +3 \sin\alpha \cos\alpha \\
& \left<\Gamma_{mix,1}(b)|J_{z}|\Gamma_{mix,2}(b)\right> = -3 \sin\alpha \cos\alpha \\
& \left<\Gamma_{mix,2}(b)|J_{z}|\Gamma_{mix,1}(b)\right> = -3 \sin\alpha \cos\alpha \\
& \left<\Gamma_{mix,1}(a)|J_{z}|\Gamma_{mix,2}(b)\right> = \left<\Gamma_{mix,2}(b)|J_{z}|\Gamma_{mix,1}(a)\right> = 0 \\
& \left<\Gamma_{mix,1}(b)|J_{z}|\Gamma_{mix,2}(a)\right> = \left<\Gamma_{mix,2}(a)|J_{z}|\Gamma_{mix,1}(b)\right> = 0
\end{align*}
\begin{align*}
& \left<\Gamma_{3/2}|J_{z}|\Gamma_{mix,2}\right>: \\
& \left<\Gamma_{3/2}(a)|J_{z}|\Gamma_{mix,2}(a)\right> = \left<\Gamma_{mix,2}(a)|J_{z}|\Gamma_{3/2}(a)\right> = 0 \\
& \left<\Gamma_{3/2}(a)|J_{z}|\Gamma_{mix,2}(b)\right> = \left<\Gamma_{mix,2}(b)|J_{z}|\Gamma_{3/2}(a)\right> = 0 \\
& \left<\Gamma_{3/2}(b)|J_{z}|\Gamma_{mix,2}(a)\right> = \left<\Gamma_{mix,2}(a)|J_{z}|\Gamma_{3/2}(b)\right> = 0 \\
& \left<\Gamma_{3/2}(b)|J_{z}|\Gamma_{mix,2}(b)\right> = \left<\Gamma_{mix,2}(b)|J_{z}|\Gamma_{3/2}(b)\right> = 0 \\
\end{align*}
For field perpendicular to the $z$-axis:
\begin{align*}
& \left<\Gamma_{mix,1}|J_{x}|\Gamma_{3/2}\right>: \\
& \left<\Gamma_{mix,1}(a)|J_{x}|\Gamma_{3/2}(a)\right> = +\frac{\sqrt{5}}{2} \cos\alpha \\
& \left<\Gamma_{3/2}(a)|J_{x}|\Gamma_{mix,1}(a)\right> = +\frac{\sqrt{5}}{2} \cos\alpha\\
& \left<\Gamma_{mix,1}(a)|J_{x}|\Gamma_{3/2}(b)\right> = +\sqrt{2}\sin\alpha \\
& \left<\Gamma_{3/2}(b)|J_{x}|\Gamma_{mix,1}(a)\right> = +\sqrt{2}\sin\alpha \\
& \left<\Gamma_{mix,1}(b)|J_{x}|\Gamma_{3/2}(a)\right> = +\sqrt{2}\sin\alpha \\
& \left<\Gamma_{3/2}(a)|J_{x}|\Gamma_{mix,1}(b)\right> = +\sqrt{2}\sin\alpha \\
& \left<\Gamma_{mix,1}(b)|J_{x}|\Gamma_{3/2}(b)\right> = +\frac{\sqrt{5}}{2} \cos\alpha \\
& \left<\Gamma_{3/2}(b)|J_{x}|\Gamma_{mix,1}(b)\right> = +\frac{\sqrt{5}}{2} \cos\alpha
\end{align*}
\begin{align*}
& \left<\Gamma_{mix,1}|J_{x}|\Gamma_{mix,2}\right>: \\
& \left<\Gamma_{mix,1}(a)|J_{x}|\Gamma_{mix,2}(a)\right> = \left<\Gamma_{mix,2}(a)|J_{x}|\Gamma_{mix,1}(a)\right> = 0 \\
& \left<\Gamma_{mix,1}(b)|J_{x}|\Gamma_{mix,2}(b)\right> = \left<\Gamma_{mix,2}(b)|J_{x}|\Gamma_{mix,1}(b)\right> = 0 \\
& \left<\Gamma_{mix,1}(a)|J_{x}|\Gamma_{mix,2}(b)\right> = -\frac{3}{2} \sin\alpha \cos\alpha \\
& \left<\Gamma_{mix,2}(b)|J_{x}|\Gamma_{mix,1}(a)\right> = -\frac{3}{2} \sin\alpha \cos\alpha \\
& \left<\Gamma_{mix,1}(b)|J_{x}|\Gamma_{mix,2}(a)\right> = -\frac{3}{2} \sin\alpha \cos\alpha \\
& \left<\Gamma_{mix,2}(a)|J_{x}|\Gamma_{mix,1}(b)\right> = -\frac{3}{2} \sin\alpha \cos\alpha
\end{align*}
\begin{align*}
& \left<\Gamma_{3/2}|J_{x}|\Gamma_{mix,2}\right>: \\
& \left<\Gamma_{3/2}(a)|J_{x}|\Gamma_{mix,2}(a)\right> = +\frac{\sqrt{5}}{2} \sin\alpha \\
& \left<\Gamma_{mix,2}(a)|J_{x}|\Gamma_{3/2}(a)\right> = +\frac{\sqrt{5}}{2} \sin\alpha \\
& \left<\Gamma_{3/2}(a)|J_{x}|\Gamma_{mix,2}(b)\right> = -\sqrt{2}\cos\alpha \\
& \left<\Gamma_{mix,2}(b)|J_{x}|\Gamma_{3/2}(a)\right> = -\sqrt{2}\cos\alpha \\
& \left<\Gamma_{3/2}(b)|J_{x}|\Gamma_{mix,2}(a)\right> = -\sqrt{2}\cos\alpha \\
& \left<\Gamma_{mix,2}(a)|J_{x}|\Gamma_{3/2}(b)\right> = -\sqrt{2}\cos\alpha \\
& \left<\Gamma_{3/2}(b)|J_{x}|\Gamma_{mix,2}(b)\right> = +\frac{\sqrt{5}}{2} \sin\alpha \\
& \left<\Gamma_{mix,2}(b)|J_{x}|\Gamma_{3/2}(b)\right> = +\frac{\sqrt{5}}{2} \sin\alpha
\end{align*}
 \section*{References}
\bibliography{PRB_Banda_resub}
\bibliographystyle{apsrev}
\end{document}